\pdfoutput=1
\documentclass{article}

\usepackage{arxiv}

\usepackage[utf8]{inputenc} % allow utf-8 input
\usepackage[T1]{fontenc}    % use 8-bit T1 fonts
\usepackage{hyperref}       % hyperlinks
\usepackage{url}            % simple URL typesetting
\usepackage{booktabs}       % professional-quality tables
\usepackage{amsfonts}       % blackboard math symbols
\usepackage{amsmath}
\usepackage{amssymb}
\usepackage{nicefrac}       % compact symbols for 1/2, etc.
\usepackage{microtype}      % microtypography
\usepackage{lipsum}
\usepackage{graphicx}
\usepackage{seqsplit}
\usepackage{float}
\usepackage{xcolor}

\newcommand{\modelname}{BioSeqVAE}
\newcommand{\modelnamesp}{BioSeqVAE }
\newcommand{\fasta}[1]{{\ttfamily\seqsplit{#1}}}

\title{How to Hallucinate Functional Proteins}

\author{
  Zak Costello\\
  Agile Biofoundry\\
  Joint BioEnergy Institute\\
  Lawrence Berkeley National Lab\\
  \texttt{zak.costello@lbl.gov*} \\
   \And
 Hector Garcia Martin \\
  Agile Biofoundry\\
  Joint BioEnergy Institute\\
  Lawrence Berkeley National Lab\\
  \texttt{hgmartin@lbl.gov} \\
}
\setcitestyle{numbers}
\begin{document}
\maketitle

\begin{abstract}

Here we present a novel approach to protein design and phenotypic inference using a generative model for protein sequences. \modelname, a variational autoencoder variant, can hallucinate syntactically valid protein sequences that are likely to fold and function. \modelnamesp is trained on the entire known protein sequence space and learns to generate valid examples of protein sequences in an unsupervised manner. The model is validated by showing that its latent feature space is useful and that it accurately reconstructs sequences. Its usefulness is demonstrated with a selection of relevant downstream design tasks. This work is intended to serve as a computational first step towards a general purpose structure free protein design tool.
\end{abstract}

\section{Introduction}

%Paragraph 1: Proteins are Very Important
Proteins are the main functional unit of life, performing a majority of tasks within the cell.  These macromolecules perform a diverse set of functions including catalysis, structural support, mechanical transduction, molecular transport, and sensing. The ability to reliably engineer proteins with a specified function in a systematic way would be transformative for biology, allowing for the explicit design of molecular machines with a targeted function for a diverse array of applications. 

However, proteins have proven difficult to engineer. Each protein is uniquely defined by a sequence of amino acids. This seemingly simple definition gives rise to dizzying complexity. While protein sequences can be trivially created by randomly choosing amino acids, not every protein sequence encodes a functional protein.  In general functional protein sequences appear to be rare in the space of all possible sequences \cite{tian_2017}. For example it has been estimated that only 1 in 1 hundred billion random protein sequences has ATP binding activity \cite{keefe2001functional}. As such, there is an underlying syntax to these sequences that is required for function to be present. Syntactic correctness gives rise to recognized secondary (e.g. alpha helices) and tertiary structures (e.g. alpha/beta-barrel domains), which in aggregate lead to function. In part due to this complexity, no general method for reliably engineering proteins exists. Engineering proteins with a desired phenotype remains a piecemeal task that requires expert level skill to perform successfully. 

%Move this paragraph to the review?
Nonetheless, the discipline of protein engineering has enabled the creation of an array of novel and useful proteins. Sustainable and inexpensive biochemicals have been developed using modern metabolic engineering practices enabled by engineered proteins \cite{siegel_2015,renata_2015,trudeau_2014}. Promising proteins for cancer therapeutics have been developed \cite{silva_2019}. Biosensors have been designed for rapid detection of various biomolecules for diagnostic and industrial purposes \cite{stein_2015}. In light of these successes, however, there is still no reliable general purpose engineering strategy.

 %Provides a path to inexpensive diagnostics in the form of biosensors \cite{stein_2015}. And is a powerful platform to develop new medicines \cite{silva_2019}. 

 %Metabolic enzymes and pathways were developed for for metabolic engineering .   Further, catalysts were designed which accelerate organic chemistry syntheses \cite{siegel_2010}. 

%Paragraph 2: Proteins are Hard to Engineer

%Paragraph 3: Generative Models offer a solution to this problem.
A recently discovered class of machine learning models, known collectively as generative models, offer a way forward. Generative models are able to take an unlabeled data set, for example pictures of human faces, and learn how to create novel examples that are semantically accurate. In the case of human faces, generated images from a trained model would have noses, mouths, and eyes in the proper locations and proportions so that they would be human recognizable. Amazingly, this is done without ever having to tell the model what any of these features are or how they are related to the concept of a face. The models are simply trained on a collection of images. Generative models have been applied successfully to many domains where unlabeled or sparsely labeled data is abundant.  Impressive results have been published where this class of models has generated realistic examples of faces \cite{karras_2017,kingma_2018}, audio \cite{huang_2018}, 3D objects \cite{NIPS2016_6096}, and RNA expression profiles \cite{lopez_2018}. 
%These models are also used to do unsupervised language translation \cite{artetxe_2018}, and design dental implants \cite{hwang_2018}. 

%Paragraph 4: Present our Model
To this end, we propose a general method for protein design and property inference which is uses only protein sequence and phenotype data. We developed a generative model variant trained on the full protein sequence space known as \modelname. \modelnamesp can generate novel protein sequences that are syntactically correct and are likely to fold and function. This model is coupled with downstream supervised models that are used to search \modelnamesp  for valid protein sequences that impart a desired function. This approach, in theory, allows for the design of a protein with any desired function. We improve over previous work in the area of using generative models on protein sequences. Our model is general in the sense that it is trained on the entire known protein sequence space and therefore can produce a protein of any type. Additionally, our model has been engineered to generate proteins up to 1000 amino acids in length. To our knowledge, 1000 amino acids is longer than any machine learning model designed to generate proteins to date. In order to contextualize our model we review the protein engineering literature generally by contrasting our model with both \emph{de novo} methods and directed evolution approaches.

%Here we propose a novel structure free approach to protein design and property inference using a deep generative model. This model is augmented by a semi-supervised approach for downstream design, classification, or regression tasks. Our objective is to build a model that intuits the underlying rules implicit in the structure of natural proteins. Then use the model that understands the syntax of protein construction as a tool to understand protein properties and to design function.

%This approach has substantial benefits distinct from both directed evolution and \emph{de novo} methods.  

%Because structure is not used to train the underlying model, much larger data sets are available for training, with over 140 million protein sequences publicly accessible on the uniprot database \cite{uniprotconsortium_2018}. This allows for the training of more accurate models than would be possible with the approximately 150 thousand structures publicly available on the protein database \cite{berman2000protein}. This model encodes proteins into a feature space which is useful for downstream tasks. [Needs Expansion] 

%[We Improve over existing generative modeling approaches by generating proteins up to 1000 amino acids in length.]  

%Paragraphs 5: Review Existing Approaches to protein Engineering
Directed evolution approaches aim to iteratively enrich for a desired function through stages of mutation and selection of an initial protein sequence.  It requires one or more starting proteins that can reasonably be evolved to have the desired function. This approach is advantageous because it does not require understanding of the relationship between sequence and function, and can still reach desired performance characteristics in a systematic way \cite{arnold_2018}. These methods have resulted in substantial success in creating industrial catalysts \cite{buller_2018,trudeau_2014}. One important limitation of these methods is that they require a protein starting point that is able to be evolved to a desired function.

\emph{De novo} methods use the principals of protein folding to design sequences with structure that results in a chosen function.  First determining the structure of a protein with the function of interest is a more reasonable task to a human designer. Then \emph{de novo} methods can find sequences that are likely to have the structure of interest. Like directed evolution, this approach has enjoyed wide success and has resulted in useful proteins designed from physical principals \cite{siegel_2010,silva_2019,siegel_2015,huang_2016,coluzza_2017}. It is distinguished from directed evolution by attempting to understand the relationship between sequence and function mediated through protein structure. Because of this, \emph{de novo} techniques are not restricted to portions of the protein sequence space that has already been explored by nature.

%Paragraph 6: Review Existing work with generative models

%\textbf{[Cite Other New Francis Arnold Paper \cite{arnold2019Feb}]} I don't think this is directly relevant to what we are trying to do... probably don't need to cite this.

%\textbf{[Cite Newish Paper on metaloprotein and peptide synthesis \cite{gupta2019feedback}  Need to read this one in detail.. so that I can compose that paragraph better.]}

There exists a small body of literature on the use generative models to infer protein properties or perform design.  To our knowledge this has only been done on a limited scale. With protein sequence data, only sets of homologous sequences have been used for training  \cite{riesselman_2018,nissen_2018,gupta2019feedback}. \modelnamesp  trains on the entire known proteome instead of specific families of proteins so that it can intuit the general syntax of a protein sequence. This more diverse training set introduces substantial challenges to constructing an well performing model. \cite{riesselman_2018,nissen_2018} used their models for downstream classification tasks and did not use them explicitly for design. Only \cite{gupta2019feedback} used their model explicitly for design. They used a novel generative modeling approach to generate small metaloproteins and peptides \cite{gupta2019feedback}.  This approach was limited in the size of proteins it could generate and did not use the entirety of the sequence space. We intend to use our model to design protein sequences with a collection of desireable protein phenotypes, instead of variants on a single family.  In a recent review, \cite{yang2018machine} postulate the possibility of a model similar to \modelname. In the next section we describe several technical hurdles which had to be overcome in order to produce a functional model.

\section{Model Development \& Methods}
\label{sec:methods}

%\textbf{[Decide which Tense to use and stick with it...]}

Applying generative models to protein sequences presents unique challenges that require special modifications to the model architecture in order to perform well. \modelnamesp is intended to be used for both phenotypic inference and design. In order to successfully perform both tasks an architechture was needed that could do all of the following: 
\begin{itemize}
\item[(i)]   handle protein sequences with variable lengths
\item[(ii)]  model interactions between distant amino acid residues
\item[(iii)]  generate novel and realistic protein sequences
\item[(iv)] encode proteins into an informative feature vector 
\end{itemize}

%Generative Model Flavors ... Transition to the choice of the Variaational autoencoder...
The first major design decision is to choose which type of generative model to use. Currently, generative models come in three flavors, variational autoencoders \cite{kingma_2013}, generative adversarial networks \cite{NIPS2014_5423}, and normalizing flows \cite{dinh_2014,dinh_2016,kingma_2018}.  All of the first three requirements above are satisfied by any generative model. Only variational autoencoders and normalizing flow based models satisfy the fourth requirement. In this work, we choose to adapt a variational autoencoder model due to their relative maturity and ease of training, but note that normalizing flow based models are worth exploring in future work. 

Generative models, such as the variational autoencoder, produce data with the same statistical properties that they were trained on.  In the case of a generative model trained on pictures of human faces, novel but semantically valid faces can be efficiently generated.  Images generated from the model have all of the characteristics required to be identified as a face including, eyes, ears, noses, and mouths. The same logic applies to \modelname, a generative model trained on functional protein sequences that fold in their native host.  Therefore, the model should produce proteins that are likely to fold and function. We verify that the model is behaving in a way consistent with this assumption. 

%A variational autoencoder can both encode protein sequences into useful feature vectors and generate valid protein examples from a provided feature vector. 
A variational autoencoder is a model that is trained to reconstruct its own input. In our case, it first encodes a protein sequence into a feature vector. The feature vector can be thought of as a summary of the important information in the protein sequence. The vector space where these feature vectors reside is generally known as the latent space. Then from that feature vector the variational autoencoder reconstructs the original protein sequence. 

A fully trained variational autoencoder can be used in two modes, either as an encoder or decoder.  The encoder can be used to take a protein sequence and find its associated feature vector.  This feature vector can then be used for downstream classification or regression tasks. For example, we can determine where a given protein is likely to be localized given its sequence. The decoder can be used to generate arbitrary sequences that are likely to fold and function by sampling from the latent space.  Further the latent space samples can be chosen so that sequences are likely to also have desired phenotypes. In the remainder of the section the model design and its uses are described detail.

\subsection{Data Sources and Processing}
%\textbf{[Question: Should I be exact here with the numbers of proteins?]}
The data to train \modelnamesp was acquired from the UniProt database \cite{uniprotconsortium_2018}. We split the UniProt sequence database into two separate parts, SwissProt and TREMBL. The SwissProt Database is hand curated and contains about 550 thousand proteins. The TREMBL part of the Database is computationally predicted and contains approximately 140 million sequences.  Sequences in the database were clustered into groups which shared over 80\% homology.  Then one sequence was chosen per cluster. This operation was performed using the Linclust command line tool \cite{steinegger_2018}. Since the goal of this model was to learn the general structures of protein sequences, we chose to include only representative sequences from clusters of proteins with similar homology. Sequences were further pruned by selecting sequences between 100 and 1000 amino acid residues in length. The data cleaning operation reduced the SwissProt and TREMBL datasets to 200 thousand and 45 million sequences respectively. For the experiments in this paper models were trained only on the SwissProt data set. The sequences were represented with one hot encoding with $21$ categories, where $20$ were amino acids and one represented sequence end.

\begin{figure}[!t]
  \centering
  \includegraphics[width=\textwidth]{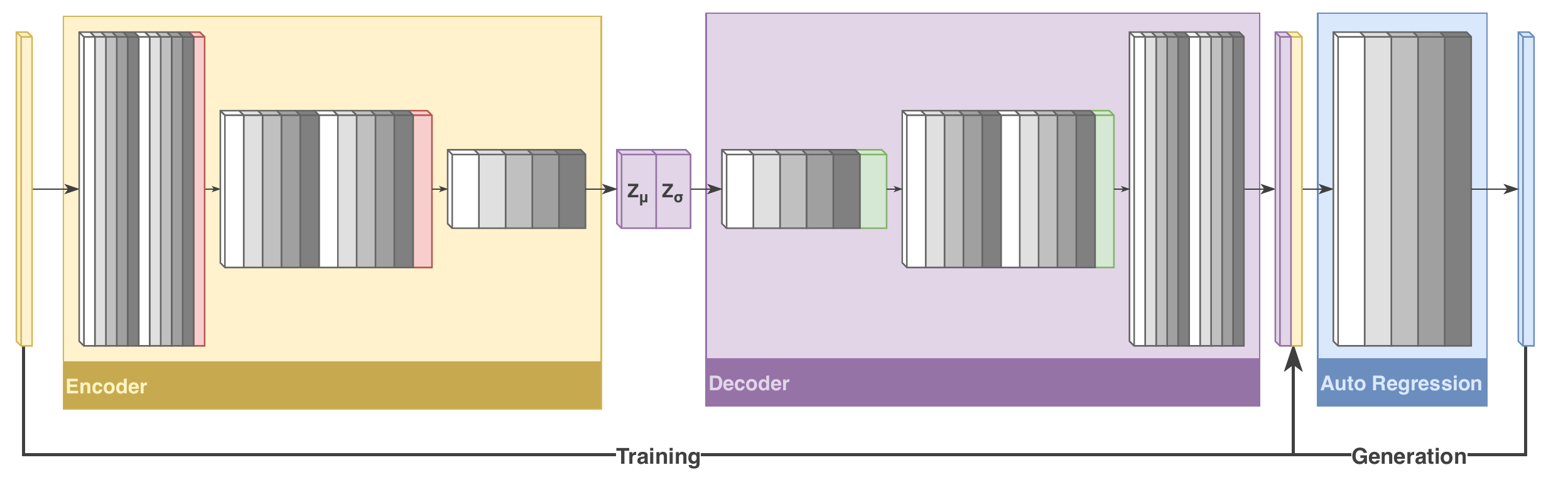}
  \caption{The architecture diagram of \modelname. Inside of each module the colored cubes each represents a layer type. Gray scale layers indicate a 1D dilated convolutional layer with skip connections in the style of resnet \cite{He_2016_CVPR}. The darkness of the grayscale layers indicates the magnitude of the dilation. Progressively darker gray indicates larger dilation. This is done in the pattern used in \cite{yu_2017}. Red layers indicate a 1D convolution where the length of the input is halved with a stride of two and the channels are doubled.  Green layers indicate the reverse operation of red via a transposed 1D strided convolution.}
  \label{fig:fig1}
\end{figure}

\subsection{Model Architecture}
%Briefly review the Variational Autoencoder
%by taking $N$ samples $x \sim X \subset \mathbb{R}^{21\times1000}$. 

Here we briefly review the variational autoencoder and discuss the relevant modifications needed to make it work on protein sequence data. For complete coverage of variational autoencoding models see \cite{kingma_2013}. We adapt a variational autoencoding model to perform unsupervised learning on protein sequences (Figure \ref{fig:fig1}). We start by constructing some data set, $\mathbf{X} = \{x_i\}_{i=1}^N$, where $\mathbf{X} \subset \mathbb{R}^{21\times1000}$. In our case this is the set of all known protein sequences after the data cleaning protocol from above. Our objective is to maximize the likelihood of our generative model $p_\theta(\mathbf{X})$ given our data. In general, this objective function,

\begin{equation}
	%\max\limit_\theta \log p_\theta(\mathbf{X}),
	\max\limits_\theta \log p_\theta(\mathbf{X}),
	\label{objective}
\end{equation}

is intractable to evaluate. The main insight of the variational autoencoder is that one can introduce a set of latent variables $z \in \mathbf{Z} \subset \mathbb{R}^m$ and break the model into an encoder and a decoder. In general $m$ is a model hyperparameter that may be chosen.  In this case we choose $m = 250$. The encoder estimates the distribution $q_\phi(z|x)$ over latent variables in $\mathbf{Z}$ given a particular datapoint $x$. The decoder provides the distribution $p_\theta(x|z)$ for output in data space $\mathbf{X}$ given a particular point in the latent space $z \in \mathbf{Z}$. Both the encoder and decoders are typically deep learning models parameterized by their weights $\theta$ and $\phi$ respectively. Starting from the objective function of the optimization problem in (\ref{objective}), we can use this insight to derive a computationally tractable lower bound on our objective using Jensen's Inequality as below:

\begin{align*} 
\log p_\theta(\mathbf{x}) &= \log \int\limits_{-\infty}^\infty p_\theta(x,z) dz\\
            %&= \log \int p_\theta(x,z)\frac{q_\phi(z|x)}{q_\phi(z|x)} dz\\
            &= \log \int p_\theta(x,z)\frac{q_\phi(z|x)}{q_\phi(z|x)} dz\\
            &= \log \mathbb{E}_{q_\phi(z|x)}\left[ \frac{p_\theta(x,z)}{q_\phi(z|x)} \right]\\
            &\geq \mathbb{E}_{q_\phi(z|x)}\left[ \log p_\theta(x,z)\right] - \mathbb{E}_{q_\phi(z|x)} \left[ \log q_\phi(z|x) \right]
\end{align*}

%Explain the Gap here... how do we get from encoders and decoders to the ELBO loss...

Now, instead of explicitly maximizing the intractable likelihood of the model, a tractable lower bound on that objective can be optimized.  This lower bound is known as the evidence lower bound [ELBO]. To gain a better intuition on this lower bound, the ELBO can be rewritten using the definition of the Kullback–Leibler [KL] divergence in an easier to interpret form,

\begin{equation*}
    \mathcal{L}_{ELBO}(\theta,\phi,x) = \mathbb{E}_{q_\phi(z|x)}\left[\log p_\theta(x|z) \right] -  D_{KL}(q_\phi(z|x)||p(z)),
\end{equation*}

where $p(z)$ is an easy to sample from distribution over $Z$.  For \modelnamesp we choose this distribution to be the standard multivariate normal distribution, $\mathcal{N}(0,I)$. The ELBO loss as expressed above has two terms with straightforward interpretations. The first term is the reconstruction loss, which measures how well a particular data point is reconstructed when run through both the encoder and decoder.  The second term represents the closeness of the latent variable distribution to the chosen simple prior distribution, $p(z)$. When this term in the loss is minimized, sampling points from the distribution of syntactically correct protein sequences efficient because valid points in $Z$ are close to the distribution $p(z)$. 

%we can simply sample points in the latent feature space from the standard normal distribution and use them to generate corresponding protein sequences in the data distribution.

%Explain the modification to our model to encode information into the latent space with flexible decoders

 For protein sequence design and phenotypic inference we need both an accurate reconstruction, and an informative latent space. To this end, we chose a high capacity decoder to encourage high reconstruction accuracy. However, this design choice can make the latent space encode uninformative features \cite{zhao_2017}. Several modifications exist to help fix this problem by constraining the amount of mutual information between $x$ and $z$ in the encoding model \cite{zhao2018information}. We use the result from \cite{zhao_2017} to augment the ELBO objective and force the model to encode informative features in the latent space.  The resulting objective has the form

\begin{equation*}
    %\mathcal{L}_{MAE}(x) = nll_loss + D_{kl}(q(z|x)||p(z))
    \mathcal{L}_{INFOVAE}(\theta,\phi,x) = \mathbb{E}_{q_\phi(z|x)}\left[\log p_\theta(x|z) \right] -  (1 - \alpha) D_{KL}(q_\phi(z|x)||p_\theta(z)) - (\alpha + \lambda - 1) D_{MMD}(q_\phi(z)||p_\theta(z)))
\end{equation*}

where $\alpha$ and $\lambda$ are hyperparameters weighting the mutual information and agreement with the chosen latent feature distribution respectively. The final term is the maximum-mean discrepancy divergence which is easily computed \cite{zhao_2017}. Now that we have motivated the model structure and chosen an objective compatible with our goals, we move forward to specific encoder and decoder design considerations.

%Encoder / Decoder Design & Features of the model
\subsection{Design of the Encoder and Decoder}
%\textbf{[Section needs to be cleaned up & Expanded with more detail]}
%\textbf{[Does talking about relationships make sense when mixed with joint distributions? Can we talk about joint distributions encoding relationships?]}
To implement a variational autoencoder a parameterized encoder, $q_\phi(z|x)$, and decoder, $p_\theta(x|z)$, must be designed. The encoder and decoder design of \modelnamesp is inspired by PixelVAE \cite{gulrajani_2016} and includes modifications made to improve its function on protein sequence data. In the particular case of encoding protein sequences, we expect the data distribution to be highly complex in the sense of having many different interactions between amino acids. Whatever model is used to estimate the joint distribution over amino acids must be sufficiently comprehensive to express every proteomic device that is known to exist. Additionally, the model must be able to capture interactions between amino acids distant in sequence space. This requirement is due to a protein sequence representing a biomolecule that is embedded in three dimensions. Additionally, we require high fidelity reconstruction and an informative latent space. These specifications are addressed by the design considerations below. 

%Because this one dimensional amino acid sequence is a representation of an object embedded in three dimensions, there may be strong relationships between amino acids distant in sequence space but nearby in distance when considered as a 3d object. 

%Model Architecture Inspiration. 
%1. Autoregressive Autoencoding \cite{gulrajani_2016,salimans_2017,oord_2016,oord_2016a,oord_2016b}. Bytenet \cite{kalchbrenner_2016}
%2. Dialated Convolutions for Increased receptive Field \cite{yu_2017,yu_2015}
%3. Forcing information in the latent space with powerful decoders. InfoVaE \cite{zhao_2017}, 

%4. Design of Resnet style layers \cite{he_2016},\cite{He_2016_CVPR}

Due to the complexity of the distribution we are trying to estimate, we start off with the assumption that the model will benefit from a very deep ResNet style convolutional network. The depth gives rise to an exponential increase in expressivity of the network \cite{raghu2017expressive}. The ResNet architecture allows the network to be efficiently trained \cite{he_2016}. The specific form of the residual layers used in both the encoder and decoder come from \cite{He_2016_CVPR}. This residual layer shows improved training times and overall performance compared to the original ResNet layer design.

In order to capture the distant interactions between residues dilated convolutions are used. Application of dilated convolutions allows for exponential instead of linear increase in the receptive field of the network as a function of depth \cite{yu_2017,yu_2015}. The chosen network architecture has a receptive field large enough to capture dependencies between any pair of amino acids in the input sequence. 

To free the variational autoencoder model from memorizing the fine details of protein sequences explicitly (e.g. the particular amino acid distribution of a beta sheet) we augment the decoder with an autoregressive module \cite{gulrajani_2016,salimans_2017,oord_2016,oord_2016a,oord_2016b}. The autoregressive module can learn the local structure of the amino acid sequence, leaving the latent space to encode the higher level details such as secondary structure into the feature space.

Combining all of the design considerations leads to the architecture visualized in Figure \ref{fig:fig1}. Inside of each module the colored cubes each represents a layer type. Gray scale layers indicate a 1D dilated convolutional layer with skip connections in the style of resnet \cite{He_2016_CVPR}. The darkness of the grayscale layers indicates the magnitude of the dilation. Progressively darker gray indicates larger dilation. This is done in the pattern used in \cite{yu_2017}. Red layers indicate a 1D convolution where the length of the input is halved with a stride of two and the channels are doubled.  Green layers indicate the reverse operation of red via a transposed 1D convolution. The encoder contains 25 convolutional resnet style blocks in total and two strided convolution layers for down scaling and channel doubling. The decoder reverses the encoder structure. In the auto regression module, gray scale blocks are causal (in the style of \cite{KalchbrennerESO16}) in addition to being 1D dilated convolution residual layers.

%Inside of each module the colored cubes each represents a layer type. Gray scale layers indicate a 1D convolutional layer with skip connections in the style of resnet \cite{He_2016_CVPR}. The darkness of the grayscale layers indicates the magnitude of the dilation. Progressively darker gray indicates larger dilation. This is done in the pattern used in \cite{yu_2017}. Red layers indicate a 1D convolution where the length of the input is halved with a stride of two and the channels are doubled. Green layers indicate the reverse operation of red via a transposed 1D convolution. In the auto regression module causal 1D convolutions are used in the style of \cite{KalchbrennerESO16}. 

\subsection{Training \& Code Availability}
To train this model the cleaned SwissProt database was used.
The model was trained end to end using the ADAM optimizer \cite{DBLP:journals/corr/KingmaB14}. Complete code will be made available coinciding with final publication.

\subsection{Protein Property Inference}
Once an instance of \modelnamesp is trained, the latent feature space can be used to predict the phenotype of a given protein.  This task is performed using a supervised learning approach. A dataset relating sequence to function must be provided in order to learn which points in latent feature space relate to specific functions. It is important to note that this can be done for any imaginable protein property for which a dataset can be gathered. Some possible properties include Gene Ontology IDs, temperature stability, EC Number, or protein localization. In practice, much of the required data has already been gathered and is readily available across many bioinformatics databases.  For this work we specifically used data present in the UniProt database. 

Supervised models are created by first using \modelnamesp to encode all protein sequences in the data set into a latent feature vector.  Then that latent feature vector and the assocated phenotype is used to train the model.  In this work, we use a random forest model from scikit-learn without parameter tuning for training. When both the unsupervised variational autoencoding model and a set of supervised phenotype models are created targeted design of function becomes possible.

\subsection{Protein Design}
\label{ProteinDesignMethod}
With \modelnamesp the design problem is reduced to a search of the latent feature space, as every point in the space is associated with a protein sequence that is likely to fold and have some function. So the design task relies on downstream models to predict how points in the latent feature space relate to desirable phenotypes. 

A set of models that relate points in the latent feature space to different phenotypes $\{f_i\}_{i=1}^N$, can be leveraged to generate enzymes with any combination desired properties. This allows design to be rephrased as an optimization problem in euclidean space as follows. 

\begin{equation}
    \label{prb:1}
    \min_z \sum\limits_{i=1}^m \alpha_i(f_i(z) - c_i)^2
\end{equation}

where $f_i$ is the $i$th model $c_i$ is the target (e.g. a specific sequence length) and $\alpha_i$ is a weight. Once solved, the optimal point in latent feature space, $\hat{z}$, is decoded to find a candidate protein to test in downstream experiments.  

\section{Results \& Discussion}

%Statement of Models Significance and section outline
\modelname 's core capability is to encode protein sequences into an information rich latent feature space and generate protein sequences that are likely to fold and function. In this section we provide evidence for these claims that motivates future experimental validation. Analyses are first performed to validate the models core function (i.e. does the model perform like a valid variational autoencoder?). Once validated, we illustrate the usefulness of \modelnamesp for design and phenotypic inference with two classification tasks. We demonstrate good performance on an enzyme classification task and a protein localization task while showing how the model can be used to design new sequences. The intent of these tasks is to demonstrate that the latent feature space encodes features that are useful for downstream learning rather than chasing state of the art performance. The ultimate objective is to develop models that allow the user to find points in the latent feature space that generate proteins with properties of interest. To emphasize this point, we generate representative sequences for each of the supervised models we present.  The section is culminated by demonstrating how sequences that are likely to have a combination of desirable properties can be generated. 

\subsection{Model Validation}

To validate the model is performing correctly, both qualitative and quantitative methods were employed. As an overall performance measure, the accuracy of encoding and then decoding the same protein was evaluated (i.e. reconstruction). Then, the distribution of the latent feature space was estimated to check that it was close a standard normal distribution. Finally, random samples were taken from the latent feature space and decoded to show qualitatively that the generated sequences look correct.

\begin{figure}[H]
  \centering
  \includegraphics[width=0.48\textwidth]{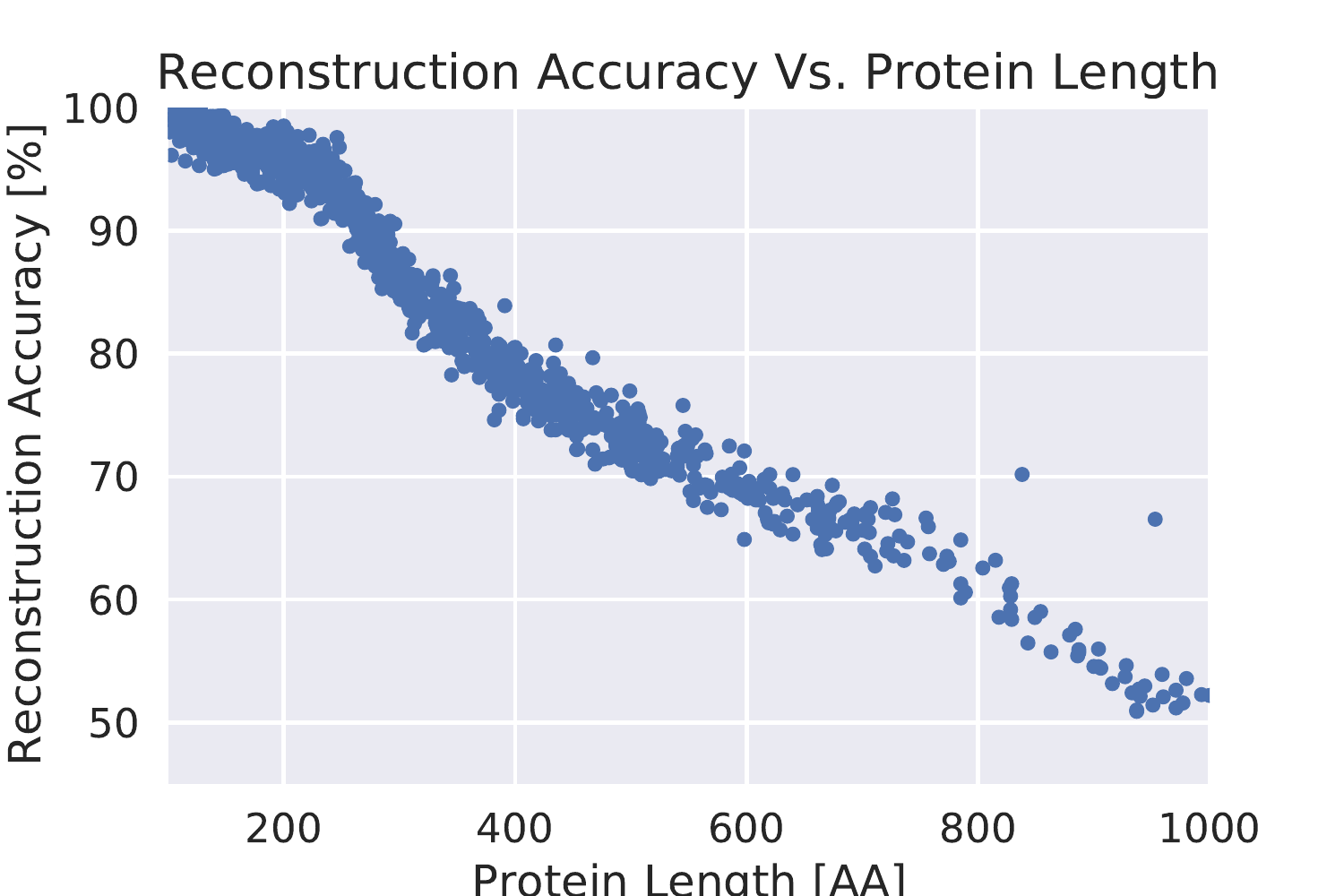}
  \includegraphics[width=0.48\textwidth]{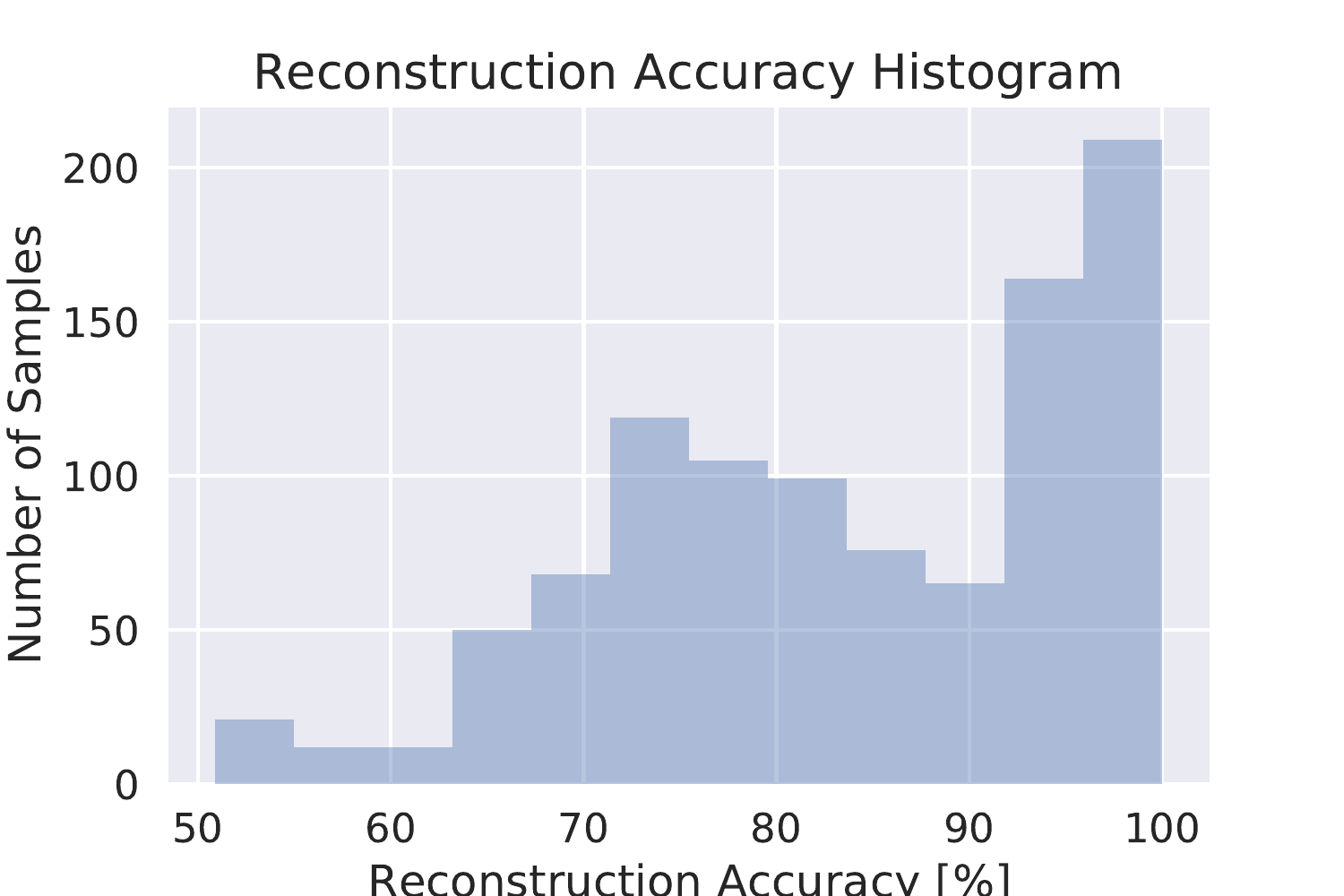}
  \caption{\modelnamesp is able to reconstruct proteins in the test set with accuracy that is dependent on length.  It is very effective at reconstructing short proteins and the accuracy trails off to around 50\% at 1000 amino acids. On average reconstruction accuracy is 83.7\% $\pm$ 12.1\%.}
  \label{fig:fig2}
\end{figure}

%Reconstruction Accuracy is Good
\modelnamesp accurately reconstructs proteins. To evaluate reconstruction accuracy, known proteins from the test set were embedded using the encoder, then decoded to predict the original sequence.  1000 random proteins from the test set were reconstructed.  The percent agreement between the actual sequence and the predicted reconstruction was calculated. The results of this test are visualized in Figure \ref{fig:fig2}. The average reconstruction accuracy is 83.7\% $\pm$ 12.1\%. The length of the protein is related to the reconstruction accuracy, with the algorithm performing better on shorter proteins. For proteins with less than 250 amino acids, we expect greater than 90\% reconstruction accuracy. We expect with increasing latent space dimension, that reconstruction accuracy will be improved.

\begin{figure}[!th]
  \begin{center}
  \includegraphics[height=5.5cm]{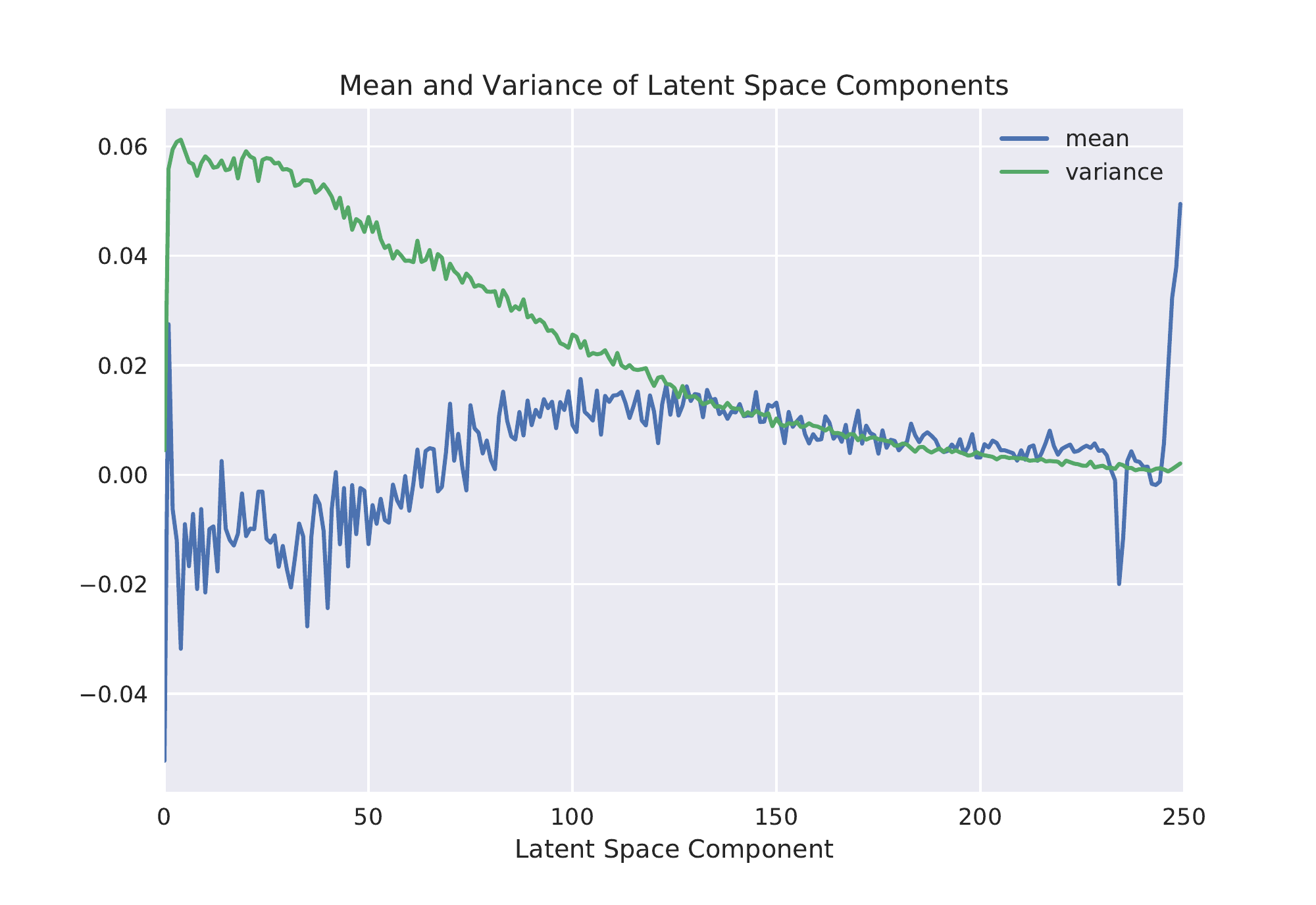}
  \includegraphics[height=5.5cm]{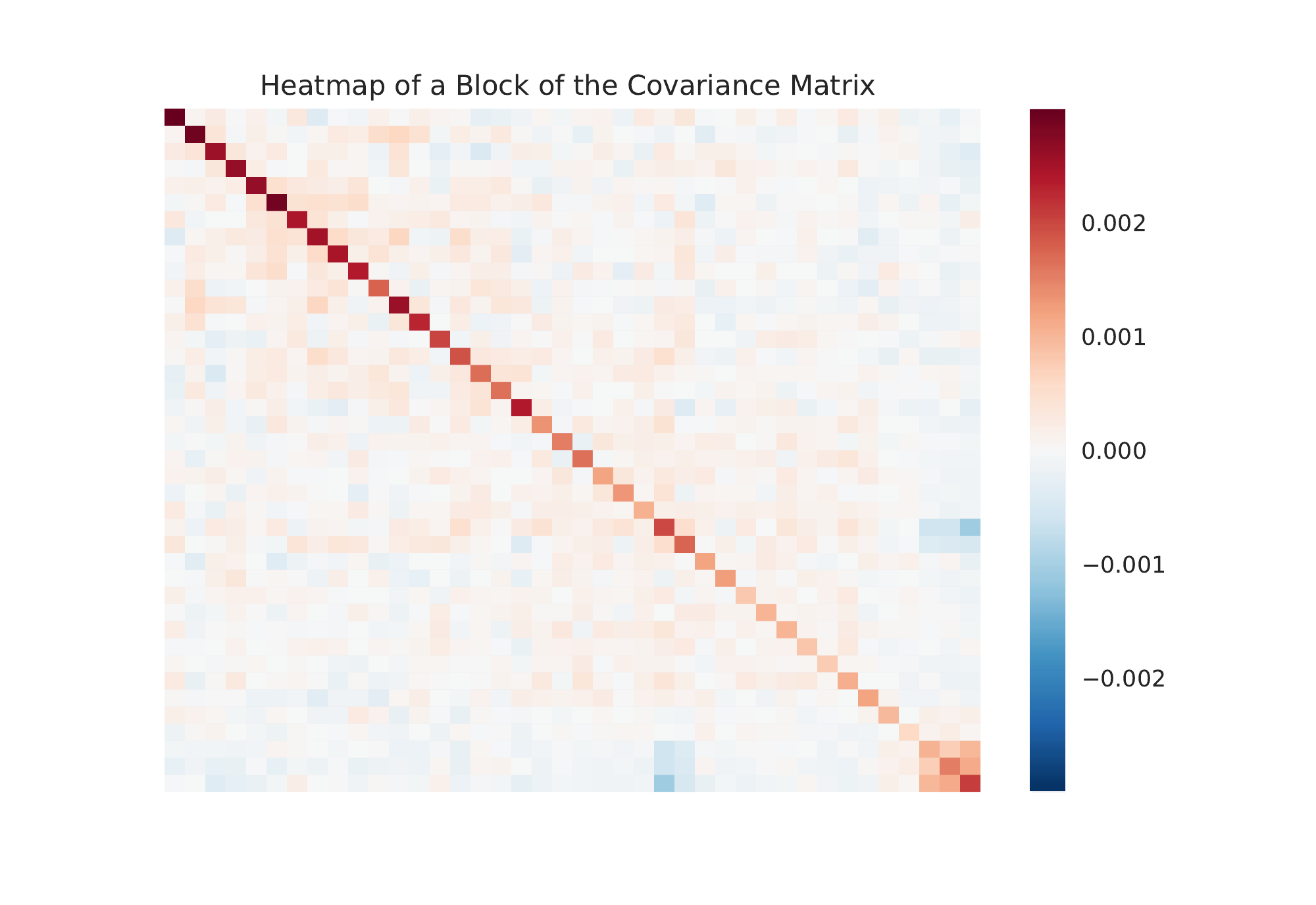}
  \end{center}
      \fasta{>{\text Random Protein 1}\\
    MAAIPEELYEAVNDASSRFVSVHEEQKSQLDLMMFSDRMVRVKSEAAHHTSMTNIEIYLKWEQMGQQSVMSVRQTSPLGLVNQFQAFATPIDAAFDRLENALRLTSLLMQGGPIDNRDRDGLLINVNYDAHGAEADGNLEAAASSASSFACPQMLDTYSGPAITKLLLQVNHLPVSPIILKADGLANLFWHIFVSMRFFTSIVHPLLLFIYYPLILGPLFEAQVPIRWPTFSVLEASYAMYHLEDPVSSLLEFSKAMALICYSCLGNSFILHDHPLHYERVAFNSGFVWGNLHLLASSL}\\
    
    \fasta{>{\text Random Protein 2}\\
    MGRLDAADVILADFGTQIVDVGAPRTKGQVEMVSVLLLHLDDPHGPIRASLGENSLDFTSPTDQLLLSPDESSVTALLLPTYLLGPVHQPAHRGGNLLLLTAAPNTRKSFPDASHTPMSHTMLDEKLKMMTREETTRDFGQRENLHEYIKNYATQYKRQTIGAVKHKNEQFESKEDWSIQQMNDGGISMFTSSAYANKSIPPGSSEAPLTESIAFLKNTAVSRAIMNPRQVNPFETIKKLEYSKKVRLNEEEPDVFNKAKLNGVKMSLNESKDSLGRPQKYPINPNAREYVNSREGLPHSLIPKHRLSFQDVGSLETHNDTMPVSLGNSIEQYAAVDAQRDDLRISEFSKDPKLFSADIDCEKEICNAMAASDLLDIWGFYAEAESKQNEGLGYILKQLPIRHLCRHSDRIIEIRGKRAPSYTVGLFASLFQCLVEFTFAPLVSTQDASSALPITQQRDEQLISVYCKVFQQQTVLEKFKQEIVWDNLKMFKDSWVTCLCVFLIPEEKKVVTTRMGYSALSNLQSRDQCFFSTLADMKIWVFPADSSRHHMKPT}\\
    
    \fasta{>{\text Random Protein 3}\\
    MTPAKKPKMSEVWDYAVGQITALSQVPEDGLPVCLGWDGGWRTSGNERVTIVELQPEAANGLAGSSTLPLQDWSWNRERDVAATQLLLRAATGAEATMSPNNLNRGKASALCLQYLTPNFTSFLAYAVSQDHALLQA}
    
  \caption{The latent space can be well aproximated as a multivariate gaussian with 250 dimensions. The dimensions of the Gaussian are close to independent. Using the mean and covariance matrix efficient samples representative of protein sequences can be synthesized. Samples of Random Protein sequences are obtained by sampling from the latent feature space then running them through the \modelnamesp decoder. Qualitatively, we can see that these proteins do not have any obvious artifacts such as long amino acid repeats. When these sequences are BLASTed against UniProt database, they have small stretches of homology with sequences that were not in the training set demonstrating that they share qualities with known proteins.}
  \label{fig:fig3}
\end{figure}

%The latent space falls on an easy to sample from distribution
The latent feature space is can be sampled from easily and produces qualitatively valid random samples. To validate that the feature space can produce good protein sequence samples, 10,000 proteins from the test set were encoded into the feature space.  The mean and covariance matrix for those encoded features was calculated. The KL divergence term in the loss encourages the latent feature space to have a standard multivariate normal distribution.  In practice we do not reach that exact distribution but approximate it. In Figure \ref{fig:fig3}, covariance matrix is plotted.  

%Random Samples taken from the distribution defined by the mean and covariance look reasonable.
Then, latent feature space samples were drawn from a multivariate normal with the estimated statistics shown in Figure \ref{fig:fig3}. One thing to notice is that the diagonal components of the covariance matrix are largest, showing that the model disentangles features from the data set into a set of features that are closer to independent. In Figure \ref{fig:fig3}, three samples from the latent feature space were reconstructed into protein sequences. Qualitatively, these sequences look good with no long stretches of amino acid repeats or other visually obvious artifacts. Additionally, the random proteins have small stretches of homology with existing proteins when BLASTed against the UniProt database.

%\textbf{[Reconstruction of GFP goes here... It is currently at 97\% reconstruction accuracy. Do I actually have to build a homology model?  What is the simplest evidence that would be good to demonstrate that reconstructions are likely to preserve function? Is 97\% homology enough?]}
%Random Samples look good
%As a qualitative measure proteins were randomly sampled from the multivariate Gaussian shown in Figure \ref{fig:fig3}. The resulting proteins, when BLASTed against the UniProt database, share small stretches of homology with known proteins that were not present in the data set. A sample of these randomly generated sequences can be found in Figure \ref{fig:fig4}.

\subsection{\modelnamesp can be used to Generate Representative Examples of Proteins with a Desired Phenotype.}

When sampling randomly from the latent feature space, the phenotype of the protein sequence that is generated is unknown. In order to discover the phenotype of that generated sequence, supervised learning methods are employed to learn the relationship between points in the latent space and a particular phenotype. This relationship will be easiest for the model to learn if \modelnamesp encodes informative features. Specifically, any set of supervised phenotype models can be used to predict which points in the latent feature space correspond to proteins of interest. For example, the models can be used to find points in the latent space which correspond to proteins that are localized in the membrane and also catalyze a desired reaction.  From those points in the latent space, \modelnamesp can hallucinate syntatically valid proteins that are likely to have the desired phenotype. In this way we can pair the strengths of several models and use them for either design or phenotypic inference. We present two examples of this as a high level proof of concept. It is worth noting that this procedure can be done for arbitrarily complex phenotypes and with any number of models.  The only requirement is that a sufficiently large supervised data set is available to train each phenotypic models. 

%Any set of supervised phenotype models, can be used to then generate proteins that have desired properties which can be predicted by those models.  This is accomplished by solving the problem phrased in (\ref{prb:1}).

%A phenotype model can be used to predict which points in the latent feature space correspond to proteins of interest. From those points in the latent space, \modelname & can hallucinate syntatically valid proteins that are likely to have the desired phenotype. In this way we can pair the strengths of two separate models and use them for either design or phenotypic inference.

\subsubsection{Enzyme Type can be Accurately Predicted and Designed using \modelname}
A powerful use case of our model is the capacity to create novel enzymes. To show an initial proof of principal, a data set of 60,000 enzyme sequences and their type were gathered. The enzyme classification scheme used was the Enzyme Commission [EC] number. The EC number is a hierarchical ontology for classifying enzymes.  It has 4 levels, where the highest level is most coarse.  This highest level is broken down into 6 general enzyme categories along with transporters as a 7th bin. We classified enzymes into each of the 6 highest level enzyme categories. Deep learning has been used to classify enzymes by EC number in the past and achieved high accuracy \cite{li2017deepre}.  Our objective was not to beat state of the art classification accuracy, but instead to show that we could generate novel examples of each of these enzyme types. Because these enzymes are hallucinated from \modelnamesp they are likely to be syntactically valid.

\begin{figure}[p]
  \begin{center}
  \includegraphics[height=5.4cm]{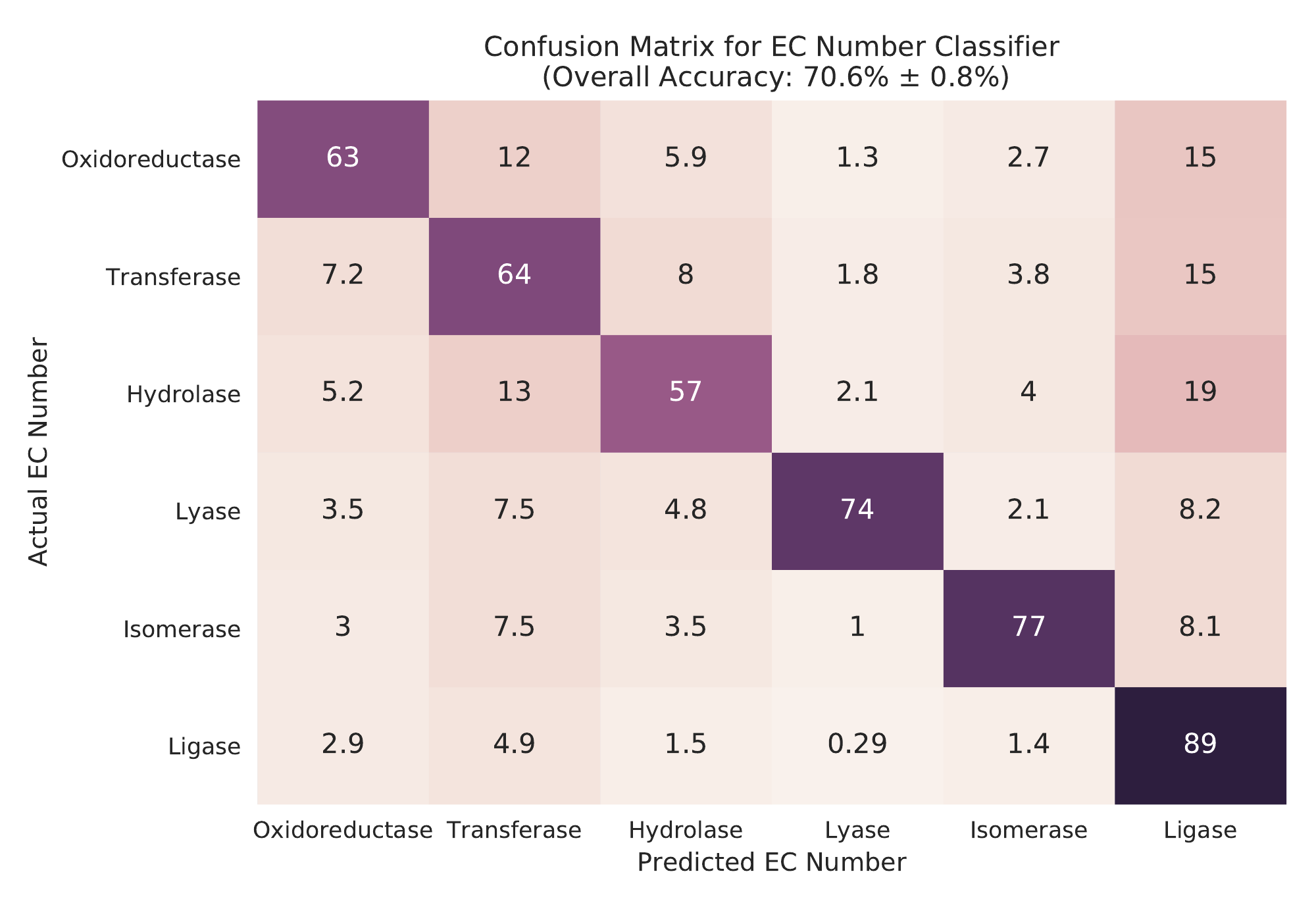}
  \includegraphics[height=5.4cm]{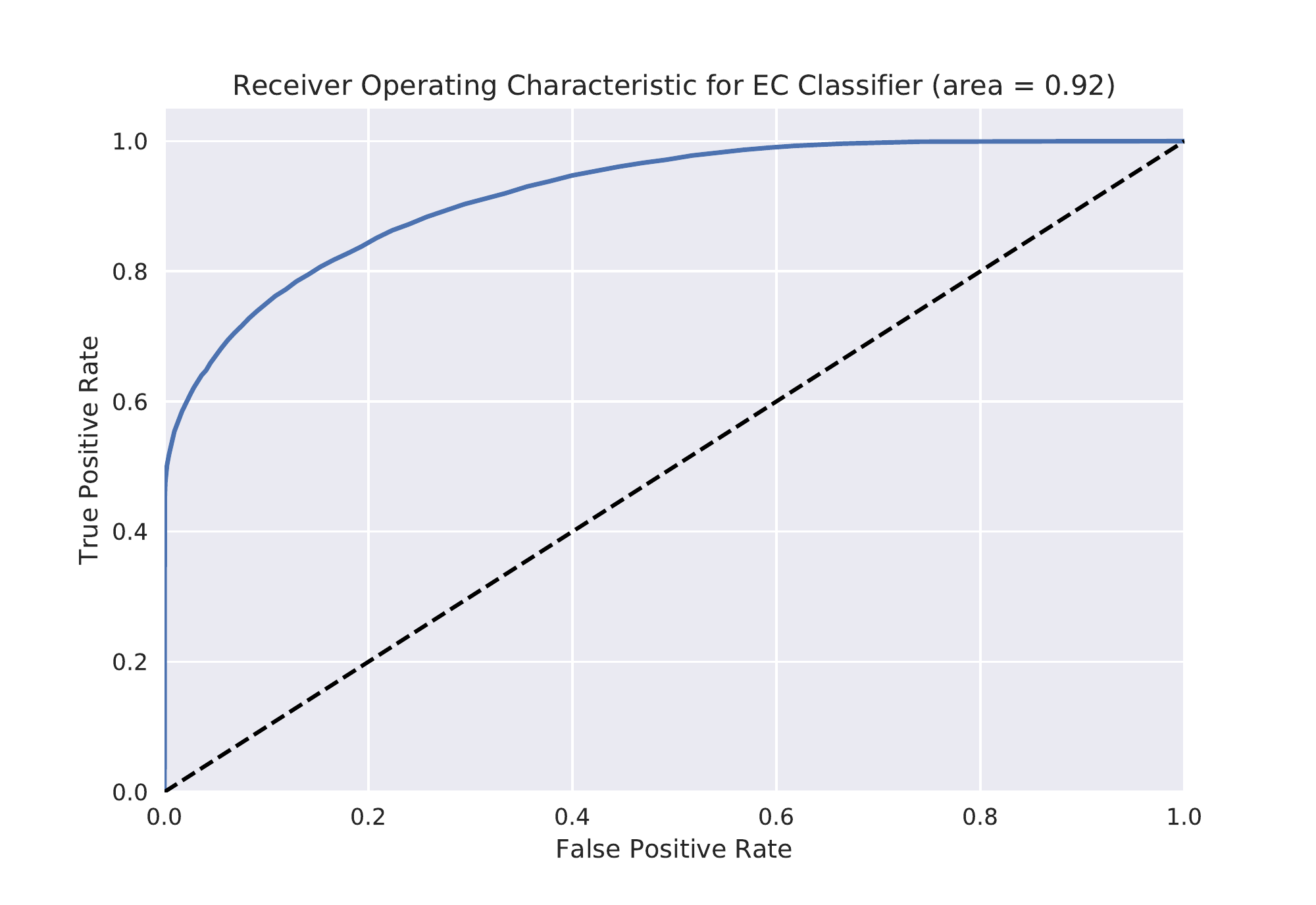}
  \end{center}
  
  \fasta{>{\text Hallucinated Oxidoreductase, Confidence 35\%} \\
    MIDPGEVTPKRAGAQKEQFGLIHRPMKPVDVALTSANQPKEFDASVKDSRGGGQRTLIRGDKPRCDWKVVRVEQEALSDILYTGTDASLQAVLDEDRRFYELAEFRKNRVRDILEDEPVSGQFFEQQDKINTGNKHTMAVAATGFDSFCMIAGAEEMIASGMPIGSARYKQQRYQGGFIEANGNESQLNGLHHLTSPVAMRCTPPMDMMAFPDDDGKQFMKGNPILPFDLGLGRKWASLTAFAGRAAARTAEGFHQGVD}\\
    
    \fasta{>{\text Hallucinated Transferase, Confidence 31\%} \\
    MSSSAGRKSTKVDYPFLLSTSCDTEYYLGMAAVFRDLDKHGRAAHDVVVKARGELAQRGILDERKSARDSFPIILLITLGPVMKEASLYPIQLIDFPLALNPEAKHAWVLHPLEHREPYGPVYPTLEAAGLPALGSVTVKLRCPAATTVEKIYIIQTGFEVAQQLNANVSTSPGMIWHARNSAPAMVVDQENILQGAPGKSTALIQTYYDSGGWIGDRFSEPKKVFHGRAAPNDNPKLLASFPLQLLMLVAVANDKSWNIEMAARGADYTAAGDAACSDVVGAATGGAIKGLPSEKRLLLNAGSTGERLATMADVLTTPGTAAMGIAADAPLYGAATGAVNDQRFFHEKVGAYPATTRAADETLTPQLQYEAGDLLSKKALAYDISAASYEACSVVFLLASRLHLAAAISGHLGAQFMELDPLSYNEAISALNFQAFHQREISAWLWRRQFLIGP}\\
    
    \fasta{>{\text Hallucinated Hydrolase, Confidence 37\%} \\
    MTASPKNRQNVYNPQFNDIEEISPVEVKSSHKIIGSHNAEVNFKNVRTDEAKQSYFIEIFENVSFYYEDGSEDAAFFEYPIKHLLKKPTSAARECGGDWLAKEEVLEFPLSTRYIEAGRDLDLQDGPLVPSVPGFARGQSPIEPNDFDEFLSFGLGITKSMHTEKSNEVGNAAFNFFKSIYDRYYGSYRRDHGSGVPAYIVRRWIPLGSGARIISRTSAIGTVISFVYSSMTYVDSEITFMGADRQAGFRARVNPLRFDIYCDARPIHKPDPSQSLHFAPDYLAEQAKITVVRRPHDQGIVYEGGLKAIVAAITFCKPFDFLSSNIYDWILKRATPVIALNDGGISGAFLLLDPHPKDDQHDRVHLKLGFAATIQLYAAEIEWAYRIQNLHEHAYFEIL}\\
    
    \fasta{>{\text Hallucinated Lyase, Confidence 26\%} \\
    MVRSEVKGFDGGRSPRRKLRRGKRGAVILIEGLVCQAVAGAVPGIARGPKGQLLAPATATASAAIAIFVLSGFYVPPGHWLTVSHHAAQAFFAVTDADNFLQRVRVRYRTQLYMLDIPRRMRMNPMAGATYLGETAADSAFENATQTGEMCAFAVVPIISLGRRSWPLSNVWIGTTVAAEVPALGLAARAWNVQIRSAAAGDLYPCYLYDTKDPPFDLYLMILAQILLDIPGQAAAVLAAIKRERLLLVQRLGTAALKA}\\
    
    \fasta{>{\text Hallucinated Isomerase, Confidence 27\%}\\
    MQATLRRQYKGPKEVVEGALLMRLAEAGFCWAGVWGRKTVVVDGRADAGVHLARILGLPEVRASEGVWAMLMLRPRLRDYLIKRIDRSPTYVQQPRLRASGAEREGQALAKSEDSAAKAPDYYKGPFDLDNHISETLEASYSKEATGHPSGHPGAPWTAPADSPGGANDRDKPAHEIMTHREDLATTPAQTFQRLEEGALVYLLLEAAALQRGQL}\\
    
    \fasta{>{\text Hallucinated Ligase, Confidence 56\%} \\
    MEKERLMYPVMHDSIQMGDAASGQRDTHMIHQGPFAFRRIRVQQEKPYYRSDDESYAISKLERPSPQISRQGDVACSTEARPPDSVFLSGAADSGTVCAKVAASGKGARNNEMKGLFGQVKELSPNAKVGLVFLKVRLAREPDSFRWTRQGGDDVALDLPRELIDRIGQTVDLLRKQPVNIPIGKERCRIDAIYQAGQYNVWQLGLVCMGCGQYFYRVKGTEAKRIYVDISLSASVTISVCEGYAHRDGMANDDTGVTSVVAIFRLPTRILDYAAARMTRQLSWPAPVDRATVDTDDDLEAILLYLLLVLNPYTYFPGPFWAVCVLRLWAGASTGMQILLGQAATDLLQYYEGMGTVYLKNNANVIIFRKLLCGMHKRYLYDI}\\
  \caption{The latent feature space can be used to both determine which enzyme class that a sequence belongs to and create novel examples of that type of enzyme. A balanced dataset of 60000 examples of enzymes with known EC numbers were used to train this model. (left) The confusion matrix shows how well each class is predicted from a given point in the latent space. (right) The combined reciever operating characteristic curve for the classifier. (bottom) Using the supervised EC number classification model and the technique outlined in section \ref{ProteinDesignMethod}, Representative sequences from each class of proteins were generated. When blasted against the UniProt database some show homology with proteins of their designated type.}
  \label{fig:fig5}
\end{figure}

A simple random forest classification model from scikit-learn \cite{scikit-learn} was applied to the collected enzyme sequence data set obtained from the UniProt database where both sequence and EC Number were known. The protein sequences were encoded into a 250 dimentional feature vector using \modelname. Then these features and the first element of the EC Number were used in a supervised learning setting to train a random forest classifier. The classifier achieved 70.6\% cross validated error (Figure \ref{fig:fig5}).  Further, it was used to generate an enzyme likely to be of each type.  

\subsection{Protein Localization can be Predicted and Designed using \modelname}
Proteins final destination within the cell is encoded in its sequence. As membrane proteins in particular can be difficult to work with in metabolic engineering applications, a method that can be used to convert a membrane protein into a cytosolic variant would be valuable. We demonstrate the generation of proteins that are likely to be secreted, localized to the membrane, the nucleus, and the cytosol. Again a phenotype model is trained to learn where a given protein is localized from its sequence.

\begin{figure}[!h]
  \begin{center}
  \includegraphics[height=5.4cm]{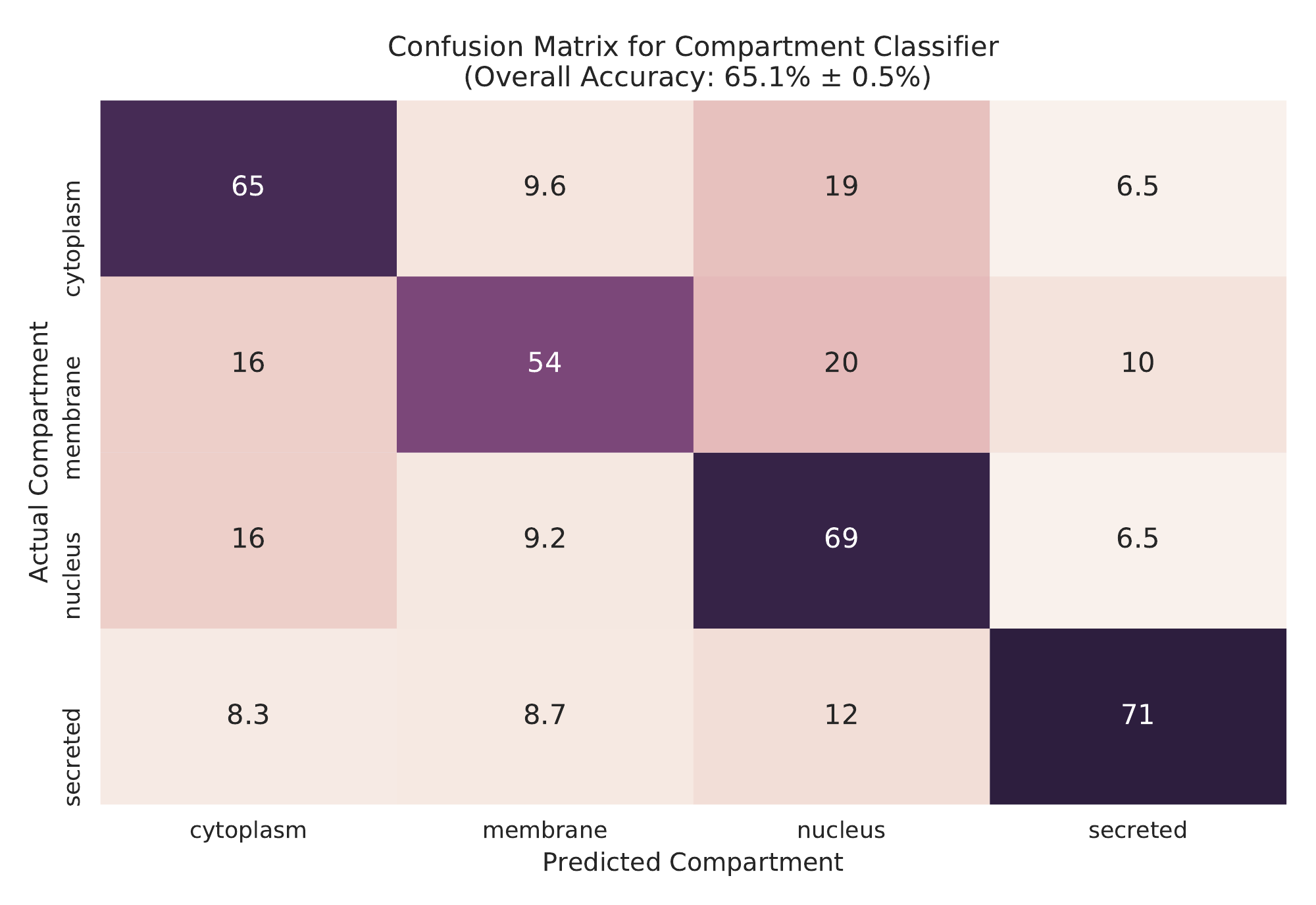}
  \includegraphics[height=5.4cm]{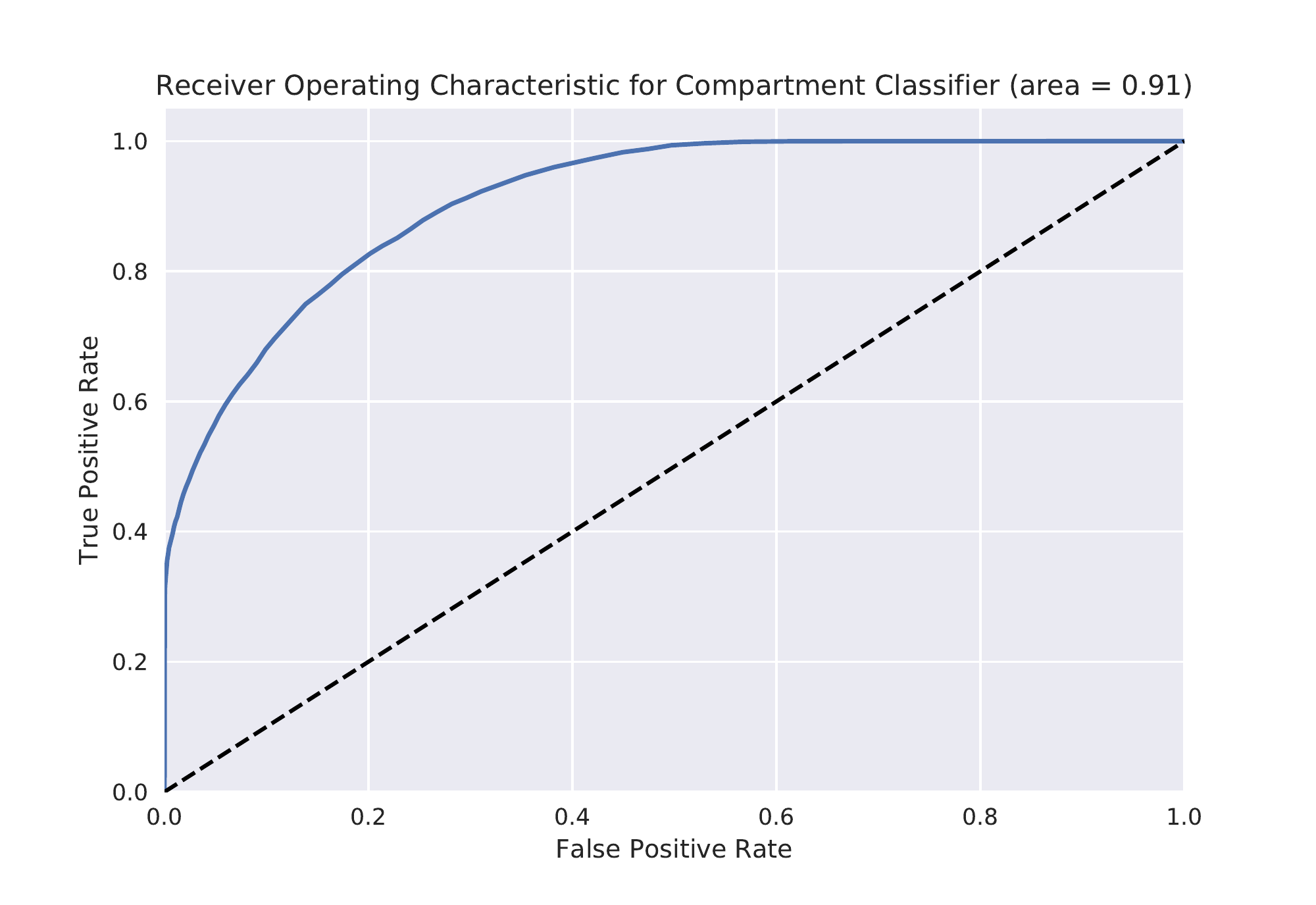}
  \end{center}
    \fasta{>{\text Hallucinated Cytoplasm Protein, Confidence 47\%}\\
    MQKQLYTGIIIEIVNLVLPNLHVTYILKACSETEIVPCAVHLDMVAGEGVSELPRTIATLSCSMTFEKYGMGRMSAGYDIPICVDAYPNSFSFLRWWDNLLDKLEGVLEIMSNLYDGFEISPYKISPAIIPRETQTEDETYDKATARGVFHVNVCYQMIQFESTGDRALMIDRLAVTAMLQSLGIMAHAFASWNFDPGMVGQVGLDGAPVGGHTFKAKHEKSSGSFDTLQAAGEIFSQWIPPIPDIHGSLSTIWWAFAAVIAAGSGFCYYLLMCARVAASIIQDRLLLFRDAYVIAGLAATTNVYPWDYFMNDTVQKAAPYAAHGLLALPVIMLIYWLLLELIYAML}\\
    
    \fasta{>{\text Hallucinated Membrane Protein, Confidence 54\%}\\
    MYASRRGSLYLRLVSQLQARDSHQRGAYSIVKYPPYTTAKLATAASIMDSKLAKVHDLRLLDVYFNNPYNEQKFHAVMQAIEIELTGCIRQGFESQGQDQNRYILNGPSGVELKGTFSGLLYIDYLYLYHVTKGHNPLDFTERRAGIHVINFFHQLDTYSAATRARAAVLHNSAANFQINHNNKIGSWLCKDCQIPSTPHHATFLGDLKERGPRMPRQALQAGARKVVELNDHNSGFICEGAHSEKATWVTASHPLDYLRKLLWHESLSSFLDAANQLLQTVGDSHKHPLLAFLLLSVSAWVLHNQLPSFRVRYNRFILLFSQLRAAPMPCECFVLKQISIKKFRLIRPRYARYAIHGGILAALPDHARKNKWVNNQEKLENGHFVAAQHDVPREAGEL}\\
    
    \fasta{>{\text Hallucinated Nucleus Protein, Confidence 76\%}\\
    MAASSHPRPQCERSWLNRGQPAETASREFFLRYGKPFLCEAPRAGVFGHCLQDQTSGQMESGGMSSVTEAAELFASGIAKWVSMRQPSVSSHFVNPLLVASWADRGLSVGKSIVTLEARYDKEVLEPVVECNRSNALEGAISPSEEYNDNDLSLNESINGKGIKELGHPTSGRAEEYLLYFPDTASKSVIVKSLSKMDVETIYCFIENPARLTSQSFTCMWTALSIQARVAAEYIGFLFLQTHYDSWDLTL}\\
    
    \fasta{>{\text Hallucinated Secreted Protein, Confidence 63\%}\\
    MLAFLLRPLLILVFAAGTSMARAGPRLPPPIGSKGSSECSSFISDCDNRVYTFEDEIRHARESAPVNSKPSEYLHRVQGHEAEQDEQFFNPASEVSACEIGAVGLMAERANVHGASVLCPAKAQYLALPIYLPFTGHTYVGAFQDERWASFCPMNTAGQVNVIYKTSDGDSQIELLIIRMAKHQSAAVVASYGSEKKLKRAQGHHTAESTNNQLISIQMIQSTGEVVGSLTTSTAAIPKYISTGLTVGRKESLTAAFAGAALEAYISATRLALAANNWYHPPFDWGKHRDDMVQL}
    
  \caption{The latent feature space can be used to both determine protein localization and create novel examples of proteins with chosen localizations. (left) The confusion matrix shows how well each class is predicted from a given point in the latent space. (right) The combined reciever operating characteristic curve for the classifier. Using the supervised compartment prediction model and the technique outlined in section \ref{ProteinDesignMethod}, Representative sequences localized to the following areas were generated: Nucleus, Membrane, Cytosol, Secreted.}
  \label{fig:fig6}
\end{figure}

The dataset was again gathered from the UniProt database. The resulting dataset contained 63,532 proteins with examples balanced evenly between each of the 4 predicted compartments.  A random forest classifier with the same hyperparameters as the previous model was applied to this dataset.  The results are visualized in Figure \ref{fig:fig6}. Average classification accuracy of the model was $65.1\%\pm 0.5\%$. From this model proteins can be found which are most likely to belong to each of the tested compartments using the method from section \ref{ProteinDesignMethod}.

%\subsubsection{Homology Between Pairs of Protein Sequences can be rapidly and accurately predicted from the latent space}
%To test the usefulness of the latent space for regression tasks, we used a random forest regression model implemented in scikit-learn to learn homology from latent space embedding.  14,000 Pairs of protein sequences were taken from the SwissProt database.  The homology percent of each pair was calculated. From that database a model was created relating both latent space embedings to the homology percent.  The Resulting model had an error standard deviation of 3.83\%. Figure \ref{fig:fig6} illustrates the cross validated performance of that model. 
%All vs All BLAST searches are computationally expensive for organizations with large sequence libraries. \modelname & may provide a way to do aproximate homology calculations by training a regression model relating the latent feature space encoding of two sequences to their percent shared homology. 

%\subsubsection{Able to Predict Temperature at which a protein functions (Data Set of Poor Quality so Temp is a bad choice...)}

%\subsubsection{Predict Protein Length? Probably less important...}

%Umap Dimensionality Reduction Technique \cite{mcinnes_2018}.

%\begin{figure}[]
%  \centering
%  \includegraphics[width=0.5\textwidth]{figures/LatentSpaceInterpolation.pdf}
%  \caption{We demonstrate interpolation in the latent space showing that the homology when interpolating in the latent space shifts from one protein to another.}
%  \label{fig:fig7}
%\end{figure}

\subsection{\modelnamesp  Can be used to Design Proteins with Multiple Properties Simultaneously}
In order to design proteins of interest, multiple phenotypes may have to be combined.  This is possible by optimizing over both of the existing classifiers simultaneously and generating a candidate sequence that is likely to be localized to the membrane and also is an oxioreductase.

\fasta{>{\text Optimal Membrane Protein Oxioreductase}\\
MLNSASTLYYPALIAALAYLLILCLIPKGVSKFSGDRSATIWAYASRGIKGARIWDNRKASVRVSFQRPHYEYYAKRLPIVVRIFKNNQWQTVDTLRKSPANGLRTCDRKGTPMEHRDTIFTVIDDYLQCQAAVMYRDYYVQPLFLYSGRMRWLKGSVSLDGRLPASPRKRRIFIGSAALAMRQDLTDGASAIAIRTGVMWAASRTPITVEGGGHLKHATQKCASLAGFAQEILLGGLVAQRPFINFNFVKRQSPTMLDDKLLDRLPWGLLAEMTHDTLALKIDCNMNETEMQYLLMIL}\\

This particular protein when blasted against the complete UniProt database has homology with a subunit of the ABC binding cassette transporter and a transmembrane protein of unknown function from \emph{Chara braunii}. While this is suggestive that the designed protein has the specified phenotype, only actual experimentation will allow the design claims to be rigorously validated.

\section{Conclusion}
%\textbf{[What Should go here? -- How to conclude...]}
We have demonstrated that realistic protein sequences can be hallucinated from an unsupervised machine learning model, \modelname. The properties of sequences can be intuited from the latent feature space of \modelname.  This paper is intended to demonstrate that protein sequences that are likely to fold and function are possible to create in the absence of structural information. This opens up the possibility to use much larger and easier to collect data sets and leverage those for the creation of novel proteins for an array of applications. This paper lays the groundwork for experimental testing of this model by synthesizing the hallucinated proteins in the lab and testing them for their designed function.  Additionally, it provides a novel way to tackle pathway completion when looking for proteins in pathways for orphaned metabolites. We fully intend to use this model to tackle these challenges.

\section{Acknowledgements}
This work was part of the DOE Agile BioFoundry (http://agilebiofoundry.org), supported by the U.S. Department of Energy, Energy Efficiency and Renewable Energy, Bioenergy Technologies Office, and the DOE Joint BioEnergy Institute (http://www.jbei.org), supported by the Office of Science, Office of Biological and
Environmental Research, through contract DE-AC02-05CH11231 between Lawrence Berkeley National Laboratory and the U. S. Department of Energy. The United States Government retains and the publisher, by accepting the article for publication, acknowledges that the United States Government retains a nonexclusive, paid-up, irrevocable, worldwide license to publish or reproduce the published form of this
manuscript, or allow others to do so, for United States Government purposes. The Department of Energy will provide public access to these results of federally sponsored research in accordance with the DOE Public Access Plan (http://energy.gov/downloads/doe-public-access-plan).

\bibliographystyle{unsrtnat}
\bibliography{references}

%\newpage

%\section*{Appendix}
%\begin{figure}[!h]
%  \centering
%  \includegraphics[width=0.48\textwidth]{figures/pshm_prediction_accuracy.pdf}
%  \includegraphics[width=0.48\textwidth]{figures/pshm_error_distribution.pdf}
%  \caption{A random forest model relates two points in the latent feature space to the homology between those two corresponding protein sequences with good accuracy. (left) The cross validated performance for each example shows that across the entire range of homology percentages prediction is consistently accurate. (right) The distribution of error is roughly gaussian with an error standard deviation of 3.83\%.}
%  \label{fig:fig6}
%\end{figure}

\end{document}